 \documentclass[final,5p,times,twocolumn]{elsarticle}

\usepackage{subfigure}
\usepackage{amssymb}
\usepackage{lipsum}
\usepackage{amsthm}
\usepackage{graphicx}
\usepackage{amsmath}
\usepackage[colorlinks=true, linkcolor=blue, citecolor=blue, urlcolor=blue]{hyperref}
\journal{Physics Letters B}

\begin{document}

\begin{frontmatter}



    \title{Interior structure of the holographic $s+p$ superconductor and chaotic-stable transition near the black hole singularity}


\author[first]{Xing-Kun Zhang}
\ead{zhangxk@nuaa.edu.cn}
\author[first]{Xin Zhao}
\ead{zhaox923@nuaa.edu.cn}
\author[second]{Zhang-Yu Nie}
\ead{niezy@kust.edu.cn}
\author[first,third]{Ya-Peng Hu}
\ead{huyp@nuaa.edu.cn}
\author[first,third]{Yu-Sen An\fnref{label1}}
\fntext[label1]{Corresponding author}
\ead{anyusen@nuaa.edu.cn}
\affiliation[first]{organization={College of Physics, Nanjing University of Aeronautics and Astronautics, Nanjing, 210016, China}}%{College of Physics, Nanjing University of Aeronautics and Astronautics, Nanjing, 210016, China}%Department and Organization
\affiliation[second]{organization={Center for Gravitation and Astrophysics, Kunming University of Science and Technology, Kunming 650500, China}}
\affiliation[third]{organization={MIIT Key Laboratory of Aerospace Information Materials and Physics,  Nanjing University of Aeronautics and Astronautics, Nanjing, 210016, China}}

\begin{abstract}
In this work, we investigate the interior structure of a holographic multi-band superconductor with the coexistence of s-wave and p-wave order parameters. %Due to the coexistence of two s-wave order parameters, the boundary system bears more interesting behaviors which can be reflected in terms of black hole interior structure. %For the region near the horizon, we find the Einstein-Rosen bridge collapse and Josephson oscillations of two scalar fields. 
Especially, we investigate the singularity structure of this multi-band model. Different from the single p-wave case, the alternation rule is jointly determined by parameters involving both s-wave order and p-wave order. In the coexistence region, we derive the Kasner alternation laws from both analytical and numerical methods which fit each other nicely. Furthermore, we find that the occurrence of the s-wave order parameter will lead to a chaotic-stable transition for the near singularity structure which matches the expectation of cosmological billiard approach. This novel transition for the near singularity structure constitutes a holographic counterpart of the secondary condensation in boundary superconducting system, offering a complementary perspective for characterizing the properties of boundary condensed matter systems.
%With the coupling between scalar field and Einstein tensor, the metric and electric field of Horndeski black hole are vastly different from the Reissner Nordstrom black hole which causes very different evaporating behaviors. In Einstein-Horndeski theory, the scalar field has a certain arbitrariness and can be regarded as an Einstein-vector theory together with Maxwell field, which has great implications for us to study the influence of the U(1) gauge field on gravity. We found that an asymptotically flat charged Einstein-Horndeski black hole will evaporate completely. But at some point, It also suggests that singularities inside the black hole do not show through, following the cosmic censorship hypothesis.
\end{abstract}




\end{frontmatter}

%\tableofcontents



\section{Introduction}
The discovery of AdS/CFT correspondence \cite{Maldacena:1997re} bridges the black hole physics and strongly coupled quantum many-body systems. The further establishment of AdS/CFT dictionary \cite{Gubser:1998bc,Witten:1998qj} provides a concise map between the two sides which makes it concrete to use gravitational tools to understand the quantum many-body system. Based on this framework, many interesting black hole solutions have been constructed to study strongly coupled quantum many-body systems. 
One most prominent example is the holographic superconductor model. The first holographic superconductor model was constructed in Ref.\cite{PhysRevD.78.065034,Hartnoll:2008kx}, where a black hole with spontaneously generated charged scalar hair can be interpreted as the gravitational dual of superconductor based on Landau-Ginzburg paradigm. The model in \cite{Hartnoll:2008kx} is an s-wave superconductor which has spatially isotropic order parameters in the electron wave function. To incorporate different symmetries of the superconducting order parameters, more kinds of holographic superconductor models have also been constructed. To name a few, by replacing the scalar field with a minimally coupled vector field, one can construct a p-wave holographic superconductor which exhibits spontaneous breaking of both the U(1) gauge symmetry and spatial rotational symmetry\cite{Gubser:2008wv,Cai:2013kaa,Cai:2013aca}. Moreover, by introducing a charged massive spin-two field propagating in the bulk, one can construct a holographic d-wave superconductor\cite{Chen:2010mk,Benini:2010pr,Kim:2013oba}. 

The above holographic superconductors only involve single order parameter. However, in many real superconducting materials, experiments show that the systems are usually characterized by multiple Fermi surfaces and superconducting energy gaps, such as $MgB_2$, iron-based superconductors and heavy fermion material \cite{CARVALHODECASTROSENE2024100083,10.1143/PTP.29.1,10.1143/PTP.36.901,Nica_2022}.Thus in order to describe these systems, multiple order parameters need to be introduced. Since then, various theoretical methods have been proposed to describe such systems \cite{PhysRevLett.3.552,PhysRevB.85.134514}. There are also many investigations from the holographic point of view. In order to understand the complex phase structures which involves multiple order parameters, researchers introduce multiple matter fields in the bulk to investigate the competition and coexistence behavior among different order parameters.  In Ref.~\cite{Cai:2013wma}, by introducing two complex scalar fields coupled to the same gauge field, an s+s coexisting phase was discovered. By incorporating different kinds of matter fields,
superconductors with different types of order parameters have also been constructed, such as s+p\cite{Nie:2013sda,Amado:2013lia,Nie:2014qma} and s+d\cite{Nishida:2014lta,Li:2014wca} holographic superconductor models.

However, most studies about holographic superconductor have only focused on the properties of black hole exterior region in the bulk while largely ignore the interior part. This is reasonable since if we only consider coarse-grained properties of the finite temperature superconducting system, the thermal mixed state is enough to characterize the system whose holographic dual only contains black hole exterior part. However, the thermal mixed state is not the only state needs to be seriously considered, instead it can be purified to be some pure state such as thermofield double state $|TFD\rangle=\frac{1}{\sqrt{Z}}\sum_{i} e^{-\beta E_{i}/2} |E_{i}\rangle_{L} |E_{i}\rangle_{R}$ \cite{Israel:1976ur,Maldacena:2001kr}. It has been established that after Lorentz evolution, the gravitational dual of the pure thermo-field double state \cite{Maldacena:2001kr} is the eternal AdS black hole (AdS wormhole) which indeed contains black hole interior part. The appearance of interior part is closely connected to the highly entangled structure of thermofield double state in the spirit of ER=EPR proposal\cite{Maldacena:2013xja}.  While the relation between black hole interior and boundary CFT remains obscure, it is widely believed that black hole interior plays pivotal role in AdS/CFT correspondence. Many researches have made a lot effort towards understanding black hole interior from different point of view, such as quantum information\cite{Stanford:2014jda,An:2022lvo,Brown:2015bva}, renormalization group flow\cite{Wang:2020nkd,Caceres:2022smh,Caceres:2022hei} and von Neumann algebra\cite{Leutheusser:2021qhd,Leutheusser:2021frk}. Black hole interior is also crucial for characterizing boundary superconductor model. During the metal-superconductor phase transition, the energy spectrum will change drastically as superconducting gap appears after transition. The change of energy spectrum certainly leads to different entangled structure of TFD state which can be reflected in terms of structure of black hole interior. After the pioneer work \cite{Hartnoll:2020fhc} , there appears various works discussing the distinct interior features for different holographic superconductor systems, such as interior of s-wave superconductor\cite{An:2022lvo,Cai:2023igv}, p-wave superconductor\cite{Cai:2021obq,Sword:2022oyg} and helical superconductor\cite{Liu:2022rsy}. It is interesting to see that while black hole exterior structure is similar for different kinds of holographic superconductor models, the interior features are vastly different. For example, while the near singularity structure of holographic s-wave superconductor with polynomial scalar potential is stable\cite{Hartnoll:2020fhc}, the near singularity structure of top-down holographic s-wave superconductor\cite{Cai:2023igv} and  holographic p-wave superconductor\cite{Cai:2021obq} is chaotic. 

%Considering the analytic continuation of the time reflection symmetric Euclidean solution for AdS black hole, the Penrose diagram can yield two asymptotic AdS regions\cite{Maldacena:2013xja}. From the perspective of one asymptotic AdS region, the other asymptotic AdS region lies behind the horizon, that is, inside the black hole. There exist two identical CFT on the two asymptotic AdS regions. When the two CFT become entangled, the dual gravitational description is that the two AdS regions are connected by an Einstein-Rosen bridge (ER = EPR)\cite{Israel:1976ur,Maldacena:2001kr}. This also indicates that the CFT on the AdS boundary is related to the internal structure of the black hole.

Since black hole interior is important in describing boundary superconductor system and multiple order parameters are ubiquitous in the studies of superconductor. It is well motivated to consider the interior features when multiple order parameters coexist. In the previous work \cite{Zhang:2025hkb}, we consider the interior structure of a holographic multi-band superconductor model which has coexisting s-wave order parameters. In this work, we are going to generalize the above work to the coexistence of s-wave and p-wave order parameters. Through investigating the interior structure of $s+p$ holographic superconductor, we find that while holographic p-wave superconductor with massless vector field exhibit chaotic singularity, including the scalar hair will destroy this chaotic structure and thus induce a chaotic-stable transition near the black hole singularity. This chaotic-to-stable transition offers a holographic description, from the black hole interior perspective, of secondary condensate formation in the boundary superconducting system.

The structure of this paper is as follows: In Sec.\ref{sec2}, we first briefly introduce the construction of holographic $s+p$ superconductor models. Sec.\ref{sec3} is the main content of this paper, we investigate the near singularity structure of this holographic superconductor model in detail and show that the singularity will transit from the chaotic type to the stable type when additional s-wave order parameter emerges. And finally in Sec.\ref{sec5}, we conclude our paper and give some outlooks.

%From Fig.\ref{FR} we can clearly see that $f(r)$ has only one real root within a reasonable range of values of $Q$, 
%\begin{equation}
%\scalebox{0.7}
%{\begin{aligned}
%     &r_{h}=\frac{1}{6} \bigg(\text{Sgn}\left(M^3-4 M Q^2\right)\sqrt{-6 \sqrt[3]{18 M^2 Q^4-64 Q^6}+9 M^2-24 Q^2}+3 M+\\&\big(\sqrt{6}\sqrt[3]{18 M^2 Q^4-64 Q^6}+\\&\sqrt{\left(3 M^2-8 Q^2\right)^2+4 \sqrt[3]{18 M^2 Q^4-64 Q^6}^2+2 \sqrt[3]{18 M^2 Q^4-64 Q^6} \left(3 M^2-8 Q^2\right)}+3 M^2-8 Q^2 \big)^{\frac{1}{2}} \bigg)
%\end{aligned}}
%\end{equation}
%where the function \text{Sgn} is the sign function. %Due to the existence of the sign function, we will divide the charge-mass ratio range into three parts $0<\frac{Q}{M}<\frac{1}{2}$, $\frac{Q}{M}=\frac{1}{2}$ and $\frac{1}{2}<\frac{Q}{M}<\frac{3}{4\sqrt{2}}$ to discuss. When $0<\frac{Q}{M}<\frac{1}{2}$,
%it's easy to prove that $r_1 $is always positive over the entire interval of $\frac{Q}{M}$. For $r_2$ to be positive, it must be satisfied $\sqrt{\frac{1}{2} \left(3-\sqrt{6}\right)}<\frac{Q}{M}<\frac{3}{4\sqrt{2}}$, which is contrary to the previous range. In the case $\frac{1}{2}<\frac{Q}{M}<\frac{3}{4\sqrt{2}}$, following the same analytical steps, one can prove that $r_1$ is always positive and $r_2$ is always negative. And when $\frac{Q}{M}=\frac{1}{2}$, $r_1$ is also always positive. And for $r_2$ to be positive it has to satisfy $Q/M> \frac {3}{4\sqrt{2}}$ which also violates the previous range. So we can conclude that under the above physical constraints, Einstein-Horndeski charged black holes only have one event horizon.
\section{Holographic superconductor model with coexistence of s-wave and p-wave order parameters:}\label{sec2}
In this section, we introduce a complex scalar field and a mass-less vector field which are both minimally coupled to the same $U(1)$ gauge field to construct the holographic $s+p$  superconductor model\cite{Nie:2014qma}. In four dimensional spacetime, the action is   
\begin{align}
S &= \frac{1}{2\kappa^2}\int \mathrm{d}^{4} x \sqrt{-g} \left( R - 2 \Lambda + \mathcal{L}_m \right), \\ 
\mathcal{L}_m &= -\frac{1}{4} F_{\mu \nu} F^{\mu \nu}   
- \frac{1}{2} \rho_{\mu \nu}^{\dagger} \rho^{\mu \nu}
-m_p^2\rho_\mu^\dagger\rho^\mu
-\tilde{D}_\mu ^\dagger\psi \tilde{D}^\mu \psi -m_s^2\psi^\dagger\psi
,
\end{align}
 where $R$ is the Ricci scalar and $\Lambda=-\frac{3}{L^{2}}$ is the cosmological constant of AdS$_4$. For simplicity, we will set $L$ to be unity throughout this work. The subscript "$\dagger$" means complex conjugate. 
 $F_{\mu\nu}=\nabla_\mu A_\nu-\nabla_\nu A_\mu$ is the field strength for the $U(1)$ gauge field. 
 $\rho_{\mu \nu}=\bar{D}_\mu\rho_\nu-\bar{D}_\nu\rho_\mu$ is the field strength for the vector field $\rho_\mu$ where the covariant derivative $\bar{D}_\mu$ reads $\bar{D}_\mu=\nabla _\mu-i q_p A_\mu$ which means that the vector field is minimally coupled to gauge field.  $\psi$ is a  complex scalar field with $\tilde{D}_\mu=\nabla_u-i q_s A_\mu$. $q_{p}$ and $q_{s}$ denote the charge of vector field and scalar field  respectively. 

The ansatz for the metric and matter field is chosen as follows:
 \begin{align}
ds^2 = \frac{1}{z^2} \left( - f(z) e^{-\chi(z)} dt^2 + \frac{1}{f(z)} dz^2 + e^{2 \zeta(z)} dx^2 + \frac{1}{e^{2 \zeta(z)}} dy^2 \right), \\
A_\nu dx^\nu = A_t(z) dt, \quad \psi=\psi(z),\quad \rho_\mu dx^\mu = \rho_x(z) dx.
\end{align}
Note that as we consider the fully back-reacting effect of the vector field, the metric ansatz is anisotropic which is characterized by the anisotropy function $\zeta(z)$. 

The equations of motion are given as follows.
\begin{equation}
\zeta^{\prime\prime}  =-\left(\frac{1}{z}+\frac{h^{\prime}}{h}\right)\zeta^{\prime}-\frac{1}{2}(z\rho_{x}^{\prime})^{2}\mathrm{e}^{-2\zeta}+\frac{q_p^{2}A_{t}^{2}\rho_{x}^{2}}{2z^{4}h^{2}}\mathrm{e}^{-2\zeta}-\frac{m_p^2\rho_x^2}{2 z^3 h}\mathrm{e}^{-2\zeta-\frac{\chi}{2}},\label{eqzeta}
\end{equation}
\begin{equation}
(z\psi^{\prime})^{\prime}  =-\frac{h^{\prime}}{h}z\psi^{\prime}-\frac{q_s^2A_t^2\psi}{z^{5}h^2}+\frac{m_s^2 \psi}{z^4h}\mathrm{e}^{-{\frac{\chi}{2}}}, \label{eqpsis}
\end{equation}
\begin{equation}
(z\rho_x^{\prime})^{\prime}  =\left(2\zeta^{\prime}-\frac{2}{z}-\frac{h^{\prime}}{h}\right)z\rho_x^{\prime}-\frac{q_p^2A_t^2\rho_x}{z^{5}h^2}+\frac{m_p^2\rho_x}{z^4h}\mathrm{e}^{-\frac{\chi}{2}},\label{eqpsip}
\end{equation}
\begin{equation}
\left(\mathrm{e}^{\frac{\chi}{2}}A_t^{\prime}\right)^{\prime}  =\frac{2q_p^2A_t\rho_x^2}{z^{3}h}\mathrm{e}^{-2\zeta}+\frac{2q_s^2A_t\psi^2}{z^5h},\label{eqphi} 
\end{equation}
\begin{equation}
\chi^{\prime}  =2z\zeta^{\prime2}+z \psi^{\prime2}+z^3\rho_x^{\prime 2}\mathrm{e}^{-2\zeta}+\frac{q_p^2A_t^2\rho_x^2}{z^{3}h^2}\mathrm{e}^{-2\zeta}+\frac{q_s^2A_t^2\psi^2}{z^5h^2},\label{eqch}
\end{equation}
\begin{equation}
h^{\prime} =\frac{\Lambda}{z^4}\mathrm{e}^\frac{-\chi}{2}+\frac{m_s^2 \psi^2 }{2z^4}\mathrm{e}^\frac{-\chi}{2}+\frac{m_p^2\rho_x^2}{2z^2}\mathrm{e}^{-2\zeta-\frac{\chi}{2}}+\frac14\mathrm{e}^\frac{\chi}{2}A_t^{'2},\label{eqh}
\end{equation} 
where the prime denotes the derivative with respect to $z$ and $h=\mathrm{e}^{-\frac{\chi}{2}}f/z^3$ is introduced for later convenience.

To numerically solve the above coupled equations of motion, we need to apply appropriate boundary conditions. Following the methods of Ref.~\cite{Hartnoll:2008kx} and Ref.~\cite{Cai:2013aca}, the procedures for constructing the holographic $s+p$ superconductor are as follows. Near the asymptotic AdS boundary $z\rightarrow0$, the asymptotic behavior of the functions are respectively
\begin{equation}
\begin{aligned}
    &A_t=\mu-z \rho+\ldots,\quad\rho_x=\rho_{x_-} z^{\Delta_{p-}}+\rho_{x_+} z^{\Delta_{p+}}+\ldots,
    \quad \\& \psi=\psi_{-}z^{\Delta_{s-}}+\psi_+z^{\Delta_{s+}}+\ldots,\quad 
    f=1+\ldots,\quad \\& \zeta=0+\ldots,\quad  \chi=0+\ldots,
\end{aligned}
\end{equation}
where the dots represent higher order terms of $z$. 
According to the AdS/CFT dictionary, $\mu,\rho,\{\rho_{x-},\psi_{-}\},\{\rho_{x+},\psi_{+}\}$ are interpreted respectively as chemical potential, charge density, sources and operator vacuum expectation values in dual field theory. The conformal dimensions of the scalar operator and vector operator are determined by the mass-dimension relation which read $\Delta_{s\pm}=\frac32 \pm \sqrt{\frac94+m_s^2}$ and $\Delta_{p \pm}=\frac12 \pm \sqrt{\frac14 +m_p^2}$. For simplicity, we fix $m_s^2=-2$ and $m_p^2=0$ to set the conformal dimension $\Delta_{s+}=2$ and $\Delta_{p+}=1$ in this work. To break the $U(1)$ symmetry spontaneously, we impose source-free boundary condition $\rho_{x-}=\psi_{-}=0$ and the order parameters associated to the spontaneous symmetry breaking are $\langle O_i\rangle=\{\rho_{x+},\psi_+\}$.

At the horizon $z=z_{h}$, except $f(z_h)=0$, regularity condition of the gauge field also requires that $A_t(z_h)=0$.  %In order to ensure that the functions satisfy the regular condition, the functions are finite at the horizon. 
%The asymptotic behavior of the functions at the horizon are
%\begin{align}
%\mathcal{F} = \mathcal{F}(z_h) + \mathcal{F}'(z_h)(z - z_h) + \ldots.\label{zhexpansion}
%\end{align}
To set the boundary conditions and use shooting method to solve the boundary value problem, we expand the functions $f$,$\chi$,$\rho_{x}$,$A_{t}$ and  $\psi$ and $\zeta$ near the horizon. 
By plugging the expansion into the Eq.(\ref{eqzeta}-\ref{eqh}), there are six independent parameters $\{z_h,\rho_x(z_h),\psi(z_h),A_{t}'(z_h),\chi(z_h),\zeta(z_h)\}$.
There are also three useful scaling symmetries associated to the equations of motion which read
\begin{equation}
    \rho_x \rightarrow \lambda \rho_x,\quad \zeta \rightarrow \zeta+\log(\lambda),\quad x\to\lambda^{-1}x,\quad y\to\lambda y \label{scalingxi}
\end{equation}
\begin{equation}
    A_t\rightarrow\lambda^{-1} A_t,\quad h\rightarrow\lambda^{-1} h,\quad t\to \lambda t, \quad \chi\rightarrow \chi + 2 \log(\lambda)\label{scalingchi}
\end{equation}
\begin{equation}
\begin{aligned}
    A_t\rightarrow \lambda^{-1}&A_t,\quad \rho_x\rightarrow \lambda^{-1}\rho_x,\quad h\rightarrow \lambda^{-3}h ,\\& z\rightarrow \lambda z, \quad (t,x,y)\to \lambda(t,x,y).\label{scalingz}
\end{aligned}
\end{equation}
By using the above three symmetries, we can firstly set $\{z_h=1,\zeta(z_h)=0,\chi(z_h)=0\}$ to perform numerical calculations. After solving  the coupled differential equations, we can use (\ref{scalingxi}) and (\ref{scalingchi}) to match the asymptotic conditions $\zeta(0)=0$ and $\chi(0)=0$.\footnote{Condition $\chi(0)=0$ means that the boundary time is equal to the bulk time at asymptotic infinity.} Therefore, we finally have three independent parameters $\{A_t'(z_h),\rho_x(z_h),\psi(z_h)\}$ at horizon. We invoke shooting method to solve this boundary value problem.  We use $\{\rho_x(z_h),\psi(z_h)\}$ as the shooting parameters to match the source-free boundary condition $\rho_{x-}=\psi_{-}=0$.\footnote{Newton-Raphson iterative method is used to search for the appropriate parameter values.} After solving the set of equations, we finally fix $\mu=1$ by using symmetry (\ref{scalingz}) which means that we choose the grand canonical ensemble to describe the boundary superconducting system.

% Scaling symmetry (\ref{scalingz}) can fix the horizon at $z_h=1$, and moreover we can start with $\chi(z_h)=\zeta(z_h)=0$ and then use scaling symmetry (\ref{scalingxi}) and (\ref{scalingchi}) to match the asymptotic conditions $\chi(0)=0$ and $\mu=1$.\footnote{Condition $\chi(0)=0$ means that the bounary time is equal to the bulk time at asymptotic infinity. Condition $\mu=1$ means that we choose to use grand canonical ensemble to describe boundary superconducting system. } Finally we have three independent parameters $\{   \rho_x(z_h),\psi(z_h),A_{t}'(z_h) \}$, we use two of the three independent parameters $\{   \rho_x(z_h),\psi(z_h),A_{t}'(z_h) \}$ as shooting parameters.  We use Newton-Raphson iterative method to search for the appropriate parameter values which match the source-free boundary condition. After finding the parameters, the equations of motion can be directly solved to extract the vacuum expectation values $\langle O_{i}\rangle$. 

\begin{figure*}[h]
    \centering
        \includegraphics[width=0.45\linewidth]{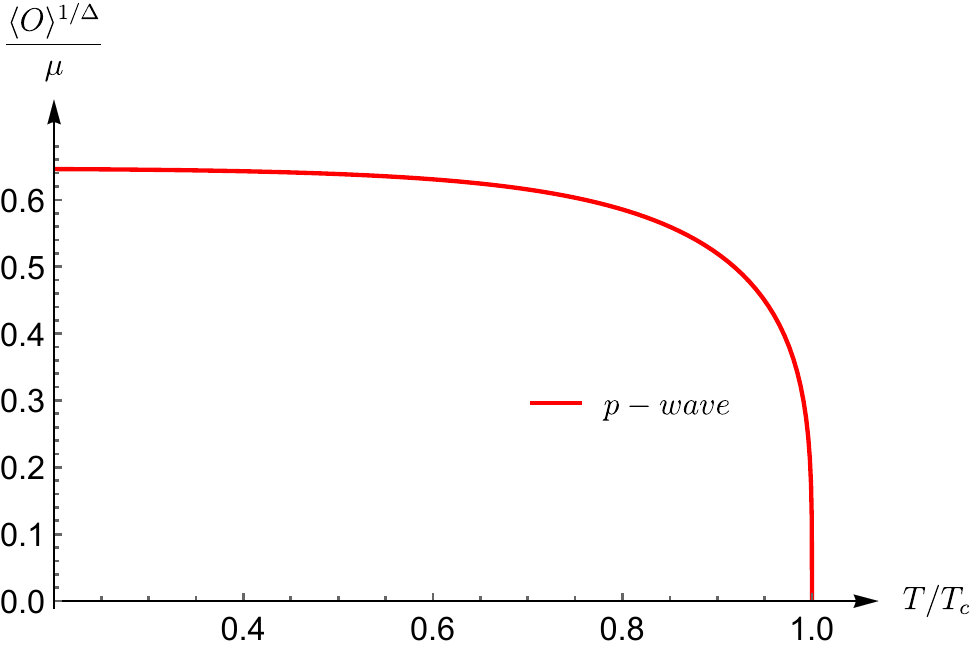}
    \includegraphics[width=0.45\linewidth]{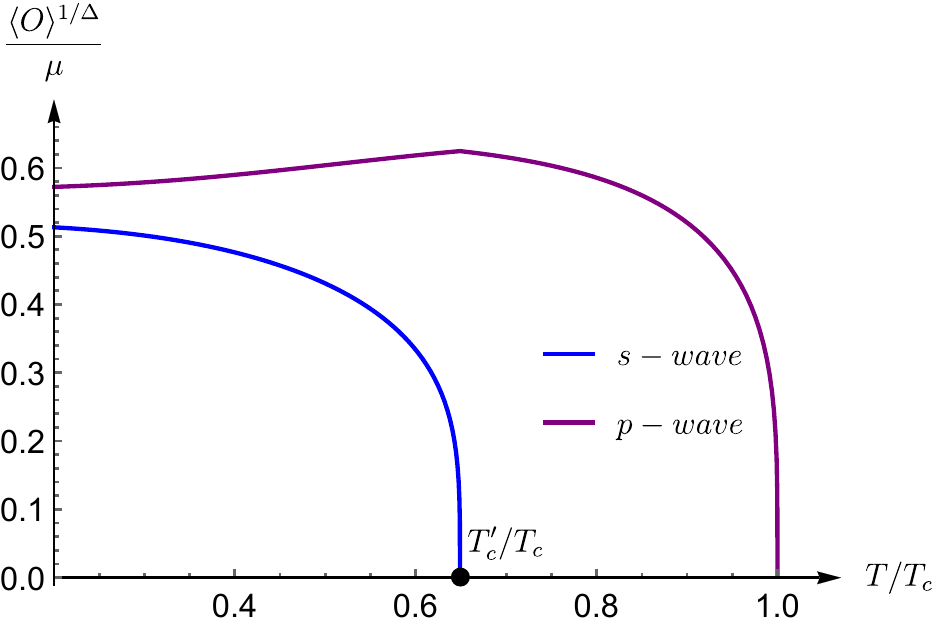}
    \caption{Phase structures in $p$ and $s+p$ holographic superconductor, we choose specific parameters which is $q_s=2.14$ and $q_p=2$. \textbf{Left panel:} Stable solution of p superconductor \textbf{Right panel:} Stable solutions of $s+p$ superconductor, where black dot ($T'_c/T_c$) denotes the critical temperature for the onset of scalar order parameter with $T'_c/T_c$=0.648942}
    \label{condensate}
\end{figure*}

We choose $q_s=2.14$ and $q_p=2$ to construct a superconductor with coexisting s-wave order and p-wave order, the condensate diagram is plotted in Fig.~\ref{condensate}.  The p-wave order in the superconductor system condenses first and then s-wave order emerges when temperature decreases further. The critical temperature $T_c$ refers to the temperature at which the single p-wave condensate first forms. 
The critical temperature of the coexisting solution when the $s$-wave order condenses on the  $p$-wave background is $T_c^\prime=0.648942T_c$. We also plot the single p-wave case for comparison in Fig.~\ref{condensate}. The configuration clearly shows that the additional s-wave order will inhibit the p-wave order.

\section{Kasner geometry and its alternation inside the $s+p$ holographic superconductor model: } \label{sec3}
\subsection{General discussion:}
After knowing the exterior region of the black hole, we can directly go beyond horizon and numerically solve the Eqs.(\ref{eqzeta}-\ref{eqh}) to get the geometry inside the black hole. By using equations of motion (\ref{eqzeta}-\ref{eqh}) or Noether theorem\cite{Cai:2021obq}, one can construct the conserved charge
\begin{equation}
    \mathcal{Q}(z) = \frac{e^{\frac{\chi}{2}}}{z^2} \left( \left(f e^{-\chi}\right)^\prime - z^2 A_t^\prime A_t + 2f \zeta^\prime e^{-\chi} \right) \label{conserve chargeeq}
\end{equation}
By using this conserved charge, it is proved in Ref.\cite{Cai:2021obq} that a smooth inner Cauchy horizon is never able to form in the presence of hair. Thus the metric function $f<0$ and $h=\mathrm{e}^{-\frac{\chi}{2}}f/z^3<0$ in the interior of the black hole. The singularity inside black hole is space-like which is distinct from the RN black hole where the singularity is time-like. Near the space-like singularity, interesting spacetime structures such as Kasner universe and mixmaster behavior can be found. Thus below, we will focus on exploring the singularity structure of the holographic $s+p$ superconuctor model. 

For the deep interior region (large $z$), the equation of motion will get vastly simplified, approximate equations of motion (\ref{eqkasner}) could be obtained by discarding higher-order infinitesimal terms \footnote{These higher-order terms can be easily shown to be vanishingly small by numerics posteriorly. }. The simplied equations take the following form

\begin{equation}
\begin{aligned}\label{eqkasner} 
 & \zeta^{\prime\prime} =-\frac{1}{z}\zeta^{\prime},~
(z\psi^{\prime})^{\prime}  =0, ~
(z\rho_x^{\prime})^{\prime}  =2\left(\zeta^{\prime}-\frac{1}{z}\right)z\rho_x^{\prime},~
\left(\mathrm{e}^{\chi/2}A_t^{\prime}\right)^{\prime}  =0, \\&
\chi^{\prime}  =2z\zeta^{\prime2}+z\psi^{\prime2},~
h^{\prime} =\frac{\mathrm{e}^{\chi/2}A_{t}^{\prime2}}{4}~.  
\end{aligned}
\end{equation} 

Solving the differential equations (\ref{eqkasner}), we can obtain analytically that
\begin{equation}
\begin{aligned}\label{eqkasnersolu}
    &\zeta = \beta \ln z + C_\zeta, \quad \psi=\sqrt{2}\gamma \ln z+C_\psi,\quad  \rho_x = C_{\rho_x} z^{2 \beta - 2},\\& A'_t = C_{A'_t}  e^{-\chi/2},\quad
\chi = 2( \beta^2+\gamma^2) \ln z + C_\chi, \quad h' = C_{h'}  e^{-\chi/2}.
\end{aligned}
\end{equation}
where $C_\zeta$, $C_\psi$, $ C_{\rho_x}$, $C_{A'_t}$, $C_\chi$, $C_{h'}$, $\beta$ and $\gamma$ are all constants.
From Eq.(\ref{eqkasnersolu}), the metric is
\begin{align}
    ds^2=C_t z^{1-\gamma^2-\beta^2}dt^2-C_zz^{-5-\gamma^2-\beta^2}dz^2+C_x z^{2\beta-2}dx^2+C_y z^{-2(1+\beta)}dy^2~.
\end{align}
Converting to the proper time $\tau \sim z^{\frac12(-3-\gamma^2-\beta^2)}$ results in an elegant metric which reads
\begin{align}
    ds^2=-d\tau^2+c_t \tau^{2p_t}dt^2+c_x\tau^{2p_x}dx^2+c_y \tau^{2p_y}dy^2~, \label{kasner metric1}
\end{align}
where $c_t$, $c_x$, $c_y$ are constants,  the metric takes the form of simple anisotropic universe which is so called "Kasner epoch". The Kasner exponent $p_{t},p_{x},p_{y}$ is the function of parameters $\beta$ and $\gamma$
\begin{align}
    p_t=\frac{-1+\gamma^2+\beta^2}{3+\gamma^2+\beta^2},~p_x=\frac{2-2\beta}{3+\gamma^2+\beta^2},~p_y=\frac{2(1+\beta)}{3+\gamma^2+\beta^2}.
\end{align}
By defining $p_\psi$ as $\psi\sim-p_\psi  \ln \tau$, it can be easily calculated that $ p_\psi=\frac{2\sqrt{2}\gamma}{3+\gamma^2+\beta^2}$.
The Kasner exponent $p_t,p_x,p_y,p_\psi$ satisfying following two simple relations
\begin{align}\label{kr}
    p_t+p_x+p_y=1, \quad p_t^2+p_x^2+p_y^2+p_\psi^2=1~.
\end{align}
Different from scalar hair, the vector hair does not appear in second relation in Eq.(\ref{kr}). From equation (\ref{eqkasnersolu}), it is evident that the introduction of the scalar field $\psi$ significantly impacts the metric function $\chi$. Compared to Ref.~\cite{Cai:2024ltu}, the interior structure of the black hole now depends on multiple parameters $\beta,\gamma$ instead of single parameter $\beta$ as shown in ~\cite{Cai:2024ltu}. We will show that in the presence of additional free parameters, the Kasner alternation rule will be very different which leads to distinct near singularity structures.  %Compared with Ref.~\cite{Zhang:2024bmf}, $\zeta$ and scalar fields play the same role. 
\subsection{Generalized Kasner Inversion behavior in the presence of two free parameters}

From the Eq.~(\ref{eqkasnersolu}), the equation of $A_{t}$ and $h$ can be easily integrated as 
\begin{equation}
    A_{t} \sim z^{1-\beta^{2}-\gamma^{2}}, \quad h\sim z^{1-\beta^{2}-\gamma^{2}}.
\end{equation}
Thus when the Kasner region satisfies $\beta^{2}+\gamma^{2}<1$ during the process towards the singularity, the electromagnetic field $A_t$ and metric function $h$ gradually accumulates and finally becomes divergent. This is inconsistent with the proof that the singularity is spacelike which indicates that the Kasner region must be unstable. This instability will cause the alternation of Kasner region. In our case, the Kasner region depends on two free parameters, thus the alternation rule will be very different from the single parameter case as explored in \cite{An:2022lvo,Cai:2023igv,Cai:2024ltu}. % Due to the limitation of the conserved charge, $h'$ must be unstable at some point.
%Fig.~\ref{figAth} 

%\begin{figure}[h]
%    \centering
%    \includegraphics[width=0.5\linewidth]{Ath.pdf}
%    \caption{Function $A_t$ and $h$ near the kasner inversion. }
%    \label{figAth}
%\end{figure}

If $h^\prime$ is non-integrable, the terms involving $h^\prime/ h$ cannot be ignored. We can get the following equations of motion 
\begin{equation}
\begin{aligned}
&\zeta^{\prime\prime} =-\left(\frac{1}{z}+\frac{h^{\prime}}{h}\right)\zeta^{\prime},\quad 
(z\psi^{\prime})^{\prime}  =-\frac{h^{\prime}}{h}z\psi^{\prime}~,
% \textcolor{red}{(z\rho_x^{\prime})^{\prime}  =2\left(\zeta^{\prime}-\frac{1}{z}-\frac{h^{\prime}}{2h}\right)z\rho_x^{\prime}-\frac{q_p^2A_t^2\rho_x}{z^{5}h^2}},\label{eqpsipinversion}\\
\left(\mathrm{e}^{\chi/2}A_t^{\prime}\right)^{\prime}  =0,~ \\&
\chi^{\prime}  =2z\zeta^{\prime2}+\frac{(z\psi^{\prime})^2}{z},~
h^{\prime} =\frac{\mathrm{e}^{-\chi/2}}{2}\left(\frac{\mathrm{e}^{\chi}A_{t}^{\prime2}}{2}\right).\label{3eq}
\end{aligned}
\end{equation}
%where 
%\begin{align}
%    h'\sim z^{-\beta^2-\gamma^2} . 
%\end{align}
%The non-integrable condition is 
%\begin{align}\label{non-integrable condition}
%    \beta^2+\gamma^2<1~.
%\end{align}
In order to solve the alternation rule analytically, we set
\begin{align}\label{def}
    \psi=\sqrt{2}\int\ \frac {\Psi}{z} dz,\quad
    \zeta=\int\frac{\Xi}{z}dz,
\end{align}
then the equation of motion (\ref{3eq}) becomes
\begin{align}
\frac{\Xi^2}{z}+\frac{\Psi^2}{z}-\frac{2\Xi'}{\Xi}+\frac{\Xi''}{\Xi'}=0~,\label{eqXI}\\
 \frac{\Xi^2}{z}+\frac{\Psi^2}{z}-\frac{2\Psi'}{\Psi}+\frac{\Psi''}{\Psi'}=0~.\label{eqPSI}
\end{align} 
The resulting differential equations (\ref{eqXI}) and (\ref{eqPSI}) are the same as those in Ref.~\cite{Zhang:2025hkb}. It can be easily seen that the two functions $\Xi(z)$ and $\Psi(z)$ satisfy the following relations
\begin{equation}\label{ev}
    2\frac{\Psi'}{\Psi}-\frac{\Psi''}{\Psi'}=2\frac{\Xi'}{\Xi}-\frac{\Xi''}{\Xi'}
\end{equation}
In order to solve the second order differential equations, we need to know the initial conditions $\Psi(z_{0}),\Psi'(z_{0}),\Xi(z_{0}), \Xi'(z_{0})$. Before alternation, the system settles down to the Kasner region which is $\psi \sim \sqrt{2} \gamma_{0} \ln z+...$ and $\zeta \sim \beta_{0} \ln z+... $. Thus based on the definition Eq.(\ref{def}), initial condition of $\Psi$ and $\Xi$ at $z=z_{0}$ must satisfy the following relations 
\begin{equation}\label{ic}
    \frac{\Xi(z_{0})}{\Psi(z_{0})}=\frac{\Xi'(z_{0})}{\Psi'(z_{0})}=\frac{\beta_{0}}{\gamma_{0}}=\lambda.
\end{equation}
Interestingly, by plugging the initial condition (\ref{ic}) into the evolution equation, we get 
\begin{equation}
    \frac{\Xi''(z_{0})}{\Psi''(z_{0})}=\frac{\beta_{0}}{\gamma_{0}}=\lambda
\end{equation}
Thus the relation of the initial condition (\ref{ic}) will be satisfied during the evolution, which means that the function $\Xi(z)$ and $\Psi(z)$ must be proportional to each other $\Xi(z)=\lambda \Psi(z)$. This makes the coupled differential equations (\ref{eqXI}) and (\ref{eqPSI}) decouple.  The decoupled equations of motion read 
\begin{equation} \label{psieq}
   (1+\lambda^{2})\Psi^2-\frac{2 z \Psi'}{\Psi}+\frac{z \Psi''}{\Psi'}=0
\end{equation}
\begin{equation} \label{xieq}
   (1+\frac{1}{\lambda^{2}})\Xi^2-\frac{2 z \Xi'}{\Xi}+\frac{z \Xi''}{\Xi'}=0
\end{equation}
The Eq.(\ref{psieq}) or Eq.(\ref{xieq}) can be directly solved , for example the solution of $\Psi$ is 
\begin{equation}
    z \Psi (\Psi-\gamma)^{\frac{1}{\gamma^{2}(1+\lambda^{2})-1}}(\frac{1}{\gamma(1+\lambda^{2})}-\Psi)^{\frac{\gamma^{2}(1+\lambda^{2})}{1-\gamma^{2}(1+\lambda^{2})}}=z_{in}
\end{equation}
where $\gamma$ and $z_{in}$ are two integration constant. From the expression of the solution, it can be easily deduced that when initially $\gamma<\frac{1}{\sqrt{1+\lambda^{2}}}$, $\Psi$ limits to $\gamma$ when $z \ll z_{in}$ and by increasing $z$ to $z \gg z_{in}$, $\Psi$ transits from $\gamma$ to $\frac{1}{\gamma(1+\lambda^{2})}$. The behavior of $\Xi$ is exactly the same as $\Psi$ which transits from $\beta$ to $\frac{\lambda^{2}}{\beta(1+\lambda^{2})}$.  
Therefore, the solution satisfies following two simple relations
\begin{align}\label{invesion}
\frac{\gamma}{\gamma_a}=\frac{\beta}{\beta_a}, \quad \gamma \gamma_a +\beta \beta_a = \frac{1}{1+\lambda^{2}}+\frac{\lambda^{2}}{1+\lambda^{2}}= 1,
\end{align}
where $\beta$ and $\gamma$ are the Kasner exponents before the Kasner alternation and $\beta_a$ and $\gamma_a$ are the exponents after 
Kasner alternation\footnote{The first condition is as a result of the proportional relation $\Xi=\lambda \Psi$.}. We see that the additional scalar field has changed the law of Kasner inversion. The condition for Kasner inversion changes from being determined solely by the parameter $\beta$ under p-wave superconductor to being jointly determined by the parameters ($\beta$, $\gamma$) in s+p model. We call this Kasner alternation as "generalized Kasner inversion".

\begin{figure*}[h]
    \centering
    \includegraphics[width=0.4\linewidth]{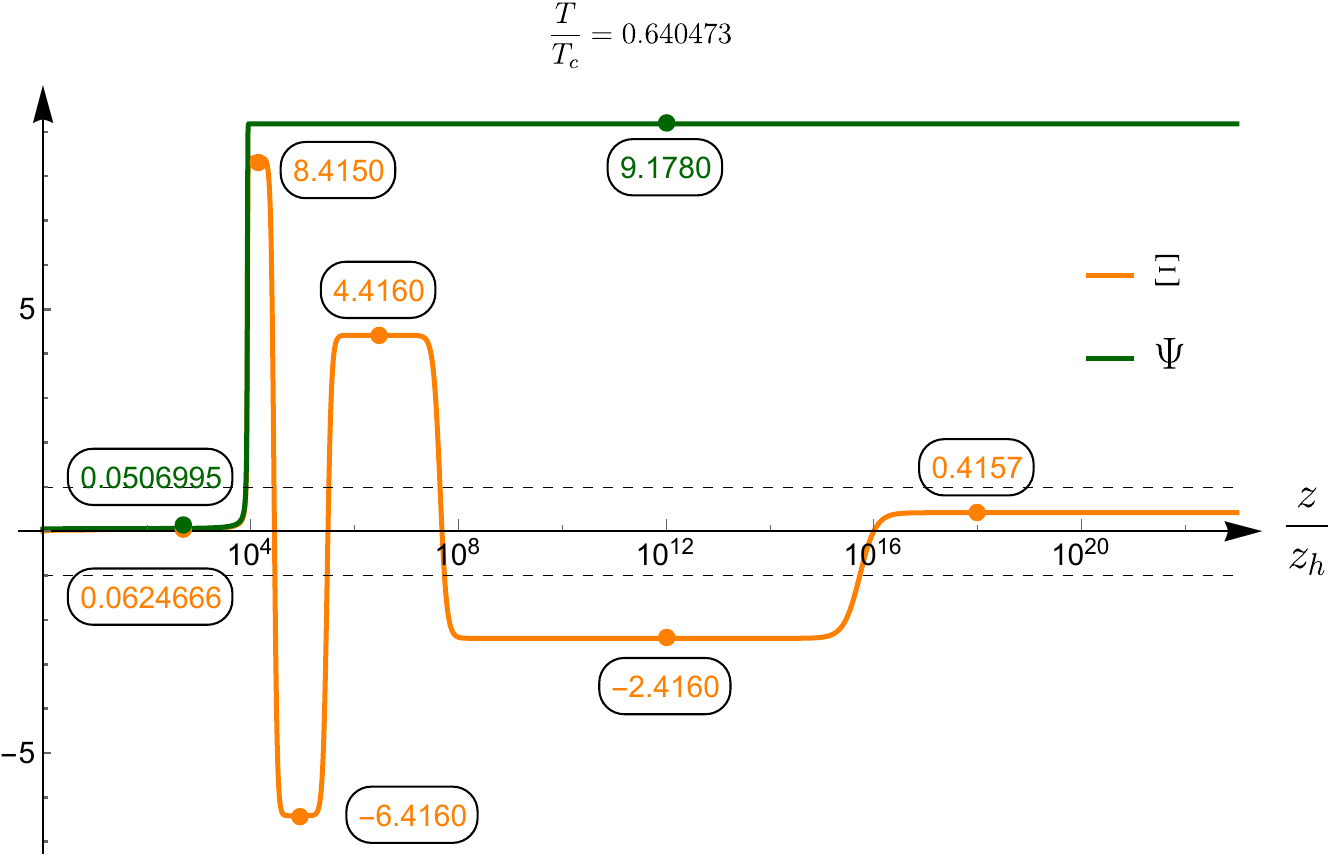}
    \includegraphics[width=0.4\linewidth]{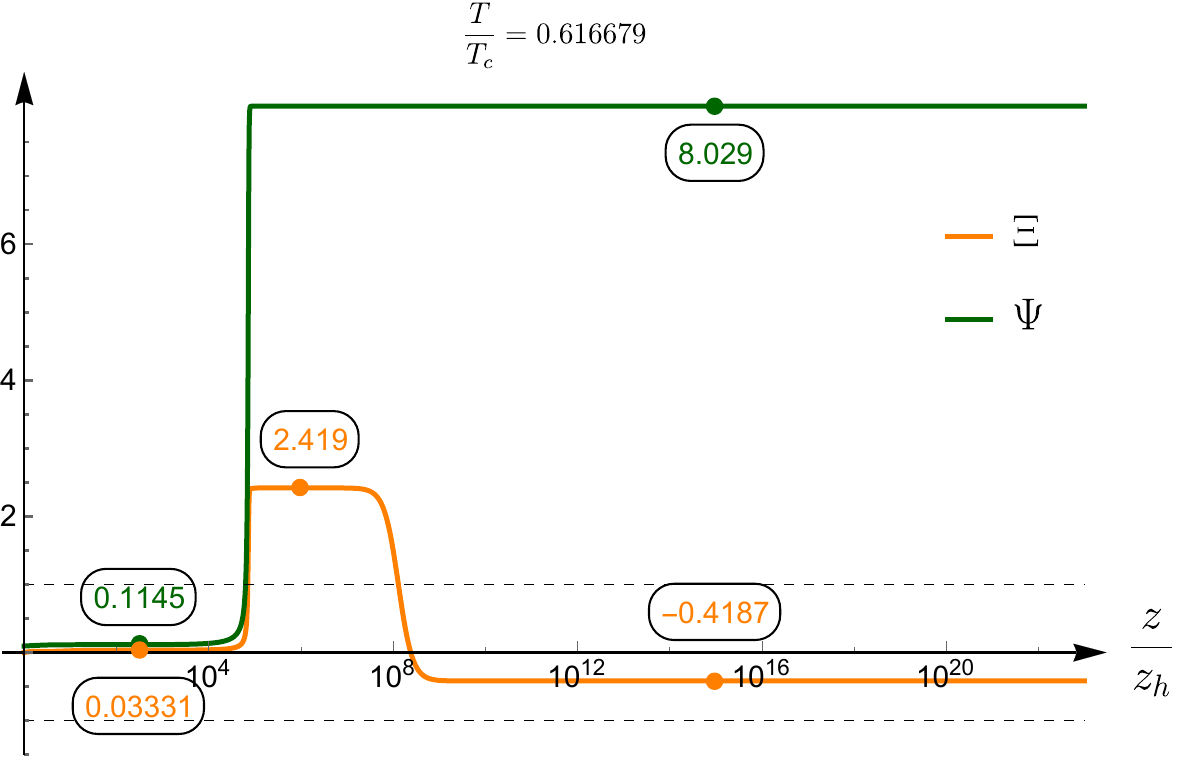}
    \includegraphics[width=0.4\linewidth]{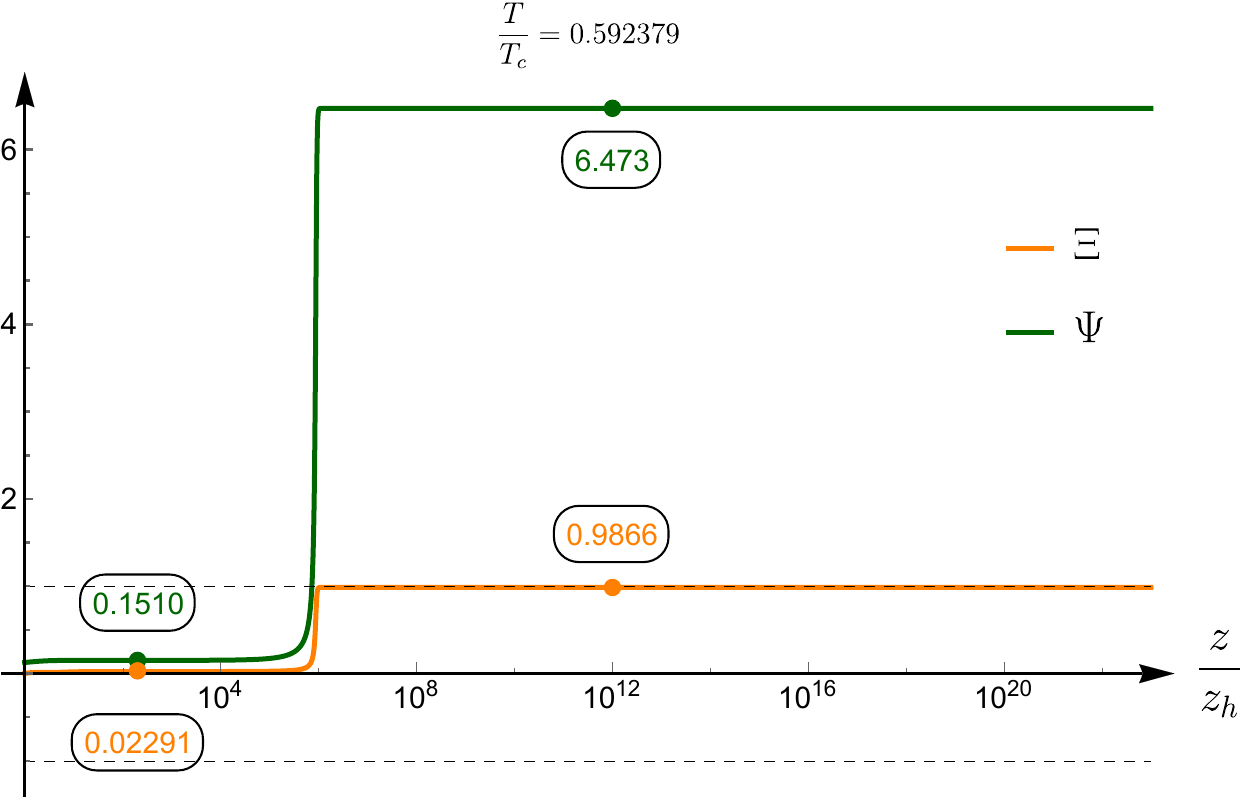}
    \includegraphics[width=0.4\linewidth]{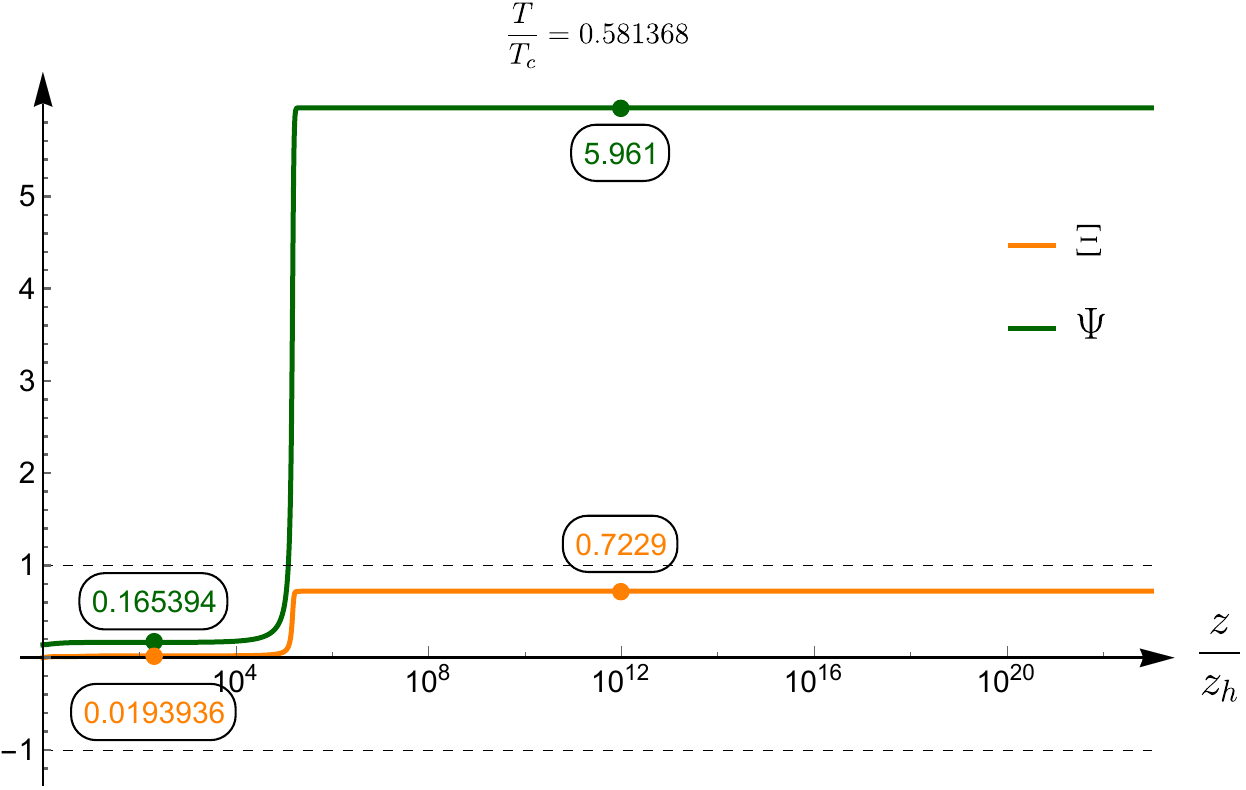}
    \caption{Kasner generalized  inversion behavior for $T=0.640473 T_c$, $T=0.616679 T_c$, $T=0.592379  T_c$ and $T=0.581368 T_c$, all of those examples satisfy the three relations in Eq.~(\ref{KasLawAll}).}
    \label{fig:kasner4}
\end{figure*}

The analytical derivations fit the numerical solutions well. In Fig.\ref{fig:kasner4}, we numerically solve the equations (\ref{eqzeta}-\ref{eqh}) to get behaviors of two fields $\Psi$ and $\Xi$ in the deep interior. We choose the temperature lying in the coexistence regions where the Kasner inversion epoch are determined by both s-wave and p-wave orders. The numerical results for the generalized Kasner inversion are also summarized in Table.\ref{table1} from which we can see that the analytical relation Eq.(\ref{invesion}) fits well with the numerical result.
%\textcolor{red}{The first two rows in Table.\ref{table1} shows the numerical results of two generalized Kasner inversion at the same temperature.}
The small error is due to the approximation we used in Eq.(\ref{3eq}) when deriving analytical result which will be smaller if the inversion occurs at larger $z$. Numerically, it is also found that the error will decrease significantly when the generalized Kasner inversion occurs at larger z region.   For example, for the case $T/T_c=0.648941$, there is actually next-round generalized Kasner inversion epoch after multiple transition and reflection period which appears at around $z\sim 10^{618}$. The full near singularity structures have been plotted in Fig.\ref{kasner fig}. For this case, the numerical error of $\gamma/\gamma_{a}-\beta/\beta_{a}$ will decrease to $O(10^{-75})$. 

\begin{table*}[h]\label{table1}
    \centering
\begin{tabular}{|c|c|c|c|c|c|c|}   
        \hline
        $T/T_{c}$& $\gamma$ & $\gamma_a$ & $\beta$ & $\beta_a$ &$\gamma \gamma_a +\beta \beta_a-1$&$\frac{\gamma}{\gamma_a}-\frac{\beta}{\beta_a}$ \\
        \hline
      0.648941& 0.000656& 0.142084& 0.063351& 15.783004& $-4\times 10^{-5}$ & $6 \times 10^{-4}$\\
        \hline
       0.637938& 0.070167& 9.336851& 0.048087& 7.171734 & $5 \times 10^{-8}$ & $8 \times 10^{-4}$\\
            \hline
       0.628536& 0.092536& 8.954179& 0.040440& 4.238670  & $-2\times10^{-8}$    &  $8\times10^{-4}$  \\
        \hline
         0.620884& 0.107204& 8.357767& 0.035618& 2.920200& $-7\times10^{-8}$    &  $6\times10^{-4}$  \\
        \hline
      0.607575& 0.129138& 7.370842& 0.028899& 1.665944  & $-3\times10^{-10}$    &  $1\times10^{-4}$ \\
         \hline   
         0.597633& 0.143723& 6.756367& 0.024814& 1.166941   &  $8\times10^{-12}$  &   $8\times10^{-6}$   \\
           \hline   
        0.586954& 0.158213& 6.208424& 0.021102& 0.841000  &  $-1\times10^{-9}$  &   $4\times10^{-4}$   \\
        \hline  
    \end{tabular}
    \caption{\label{table1} The exponents before and after the Kasner generalized  inversion calculated by numerics. It can be found that the analytical result and numerical result fit each other nicely in all cases.}
\end{table*}

\subsection{Kasner Transition and Reflection: }
If $h'$ is integrable, terms of $h'/h$ in the equations of motion (\ref{eqzeta})-(\ref{eqh}) can be ignored.
According to the integrability conditions of $\rho_x'$, the Kasner alternative law can be divided into Kasner transition and Kasner reflection\cite{Cai:2024ltu}. As can be seen in Eq.(\ref{eqzeta})-(\ref{eqh}), the additional scalar field has no influence on the process of Kasner transition and Kasner reflection if term $h'/h$ is neglected. Thus the results of Kasner transition and reflection obtained in \cite{Cai:2024ltu} are the same in holographic $s+p$ superconductor case. Therefore, if $\beta>1$, the Kasner alternation rule will be the Kasner transition where $\beta+\beta_{T}=\frac{2}{d-2}$ which is $\beta+\beta_{T}=2$ in four dimensions. If $\beta<-1$, the Kasner alternation rule is Kasner reflection where $
\beta+\beta_{R}=-2$.  

\subsection{Chaotic-stable transition of the near singularity structure}
Above results illustrate that the Kasner alternation rule in holographic $s+p$ superconductor can be summarized as follows 
\begin{equation}\label{KasLawAll}
\begin{cases}
\mathrm{Transition}: \beta+\beta_T=2, (\beta>1)\,, \\
\mathrm{Generalized \ inversion 
}: \quad \frac{\gamma}{\gamma_a}=\frac{\beta}{\beta_a}, \quad \gamma \gamma_a +\beta \beta_a =1, (\beta^{2}+\gamma^{2}<1),\\
\mathrm{Reflection}: \beta+\beta_R=-2, (\beta<-1)\,,\\
\end{cases}
\end{equation}
%Let's start from $\beta$.
%If $\beta$>1, the Kasner transition restriction condition is satisfied. The kasner transition occurs in the system.
%If $\beta$<-1, kasner reflection occurs in the system.
%These two transformation results are consistent with \cite{Cai:2024ltu}.
Compared with the Kasner alternation results for single p-wave holographic superconductor 
\begin{equation}\label{KasLawAll1}
\begin{cases}
\mathrm{Transition}: \beta+\beta_T=2,\quad \beta>1\,, \\
\mathrm{Inversion}: \quad \beta\,\beta_{I}=1,\quad -1<\beta<1\,,\\
\mathrm{Reflection}: \beta+\beta_R=-2,\quad \beta<-1\,,\\
\end{cases}
\end{equation}
we find that whether generalized Kasner inversion happens depends on both parameters $\gamma$ and $\beta$. The previous result (\ref{KasLawAll1}) indicates that there is no region for $\beta$ to make the Kasner epoch stable. However, with the addition of $\gamma$, the stable Kasner epoch can appear when $|\beta|<1$ and $\beta^{2}+\gamma^{2}>1$. 
According to the left one of generalized inversion law (\ref{invesion}), the ratio of $\beta$ and $\gamma$ before and after the occurrence of generalized Kasner inversion is equal.  If $\beta^{2}+\gamma^{2}<1$, absolute value of $\beta$ and $\gamma$ must increase after the generalized Kasner inversion because of the condition $\gamma \gamma_{a}+\beta \beta_{a}=1$. After generalized Kasner inversion, there are two possibilities. First is that the end result satisfies $\beta_{a}^{2}+\gamma_{a}^{2}>1$ while maintaining $|\beta_{a}|<1$, this directly leads to the stable Kasner region. The second is for $\gamma \ll 1$ which is the case for temperature near the transition point $T'_{c}$, for this case $\beta_{a} \approx \frac{1}{\beta}$ which leads to $|\beta_{a}|>1$, then Kasner transition or reflection will happen. After the Kasner transition or reflection, $\beta$ will finally satisfy $|\beta|<1$, and the subsequent epoch depends on the value $\beta^{2}+\gamma^{2}$. If $\beta^{2}+\gamma^{2}<1$, further generalized Kasner inversion will happen and if $\beta^{2}+\gamma^{2}>1$ the system settles down to a stable singularity. However, note that every time the system experience a generalized Kasner inversion, the absolute value of $\gamma$ will increase which must lead to condition $\beta^{2}+\gamma^{2}>1$ after multiple round of generalized Kasner inversion, thus finally the system must settle down to the stable Kasner epoch. 

\begin{figure*}[h!]
    \centering
    \includegraphics[width=0.475\linewidth]{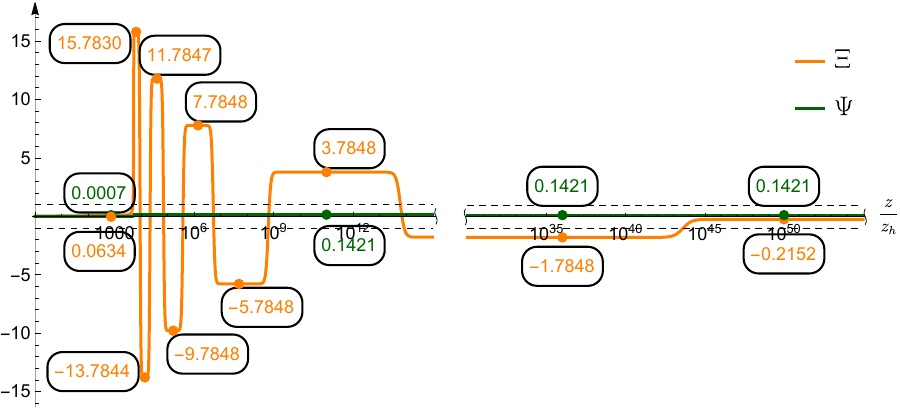}
    \includegraphics[width=0.475\linewidth]{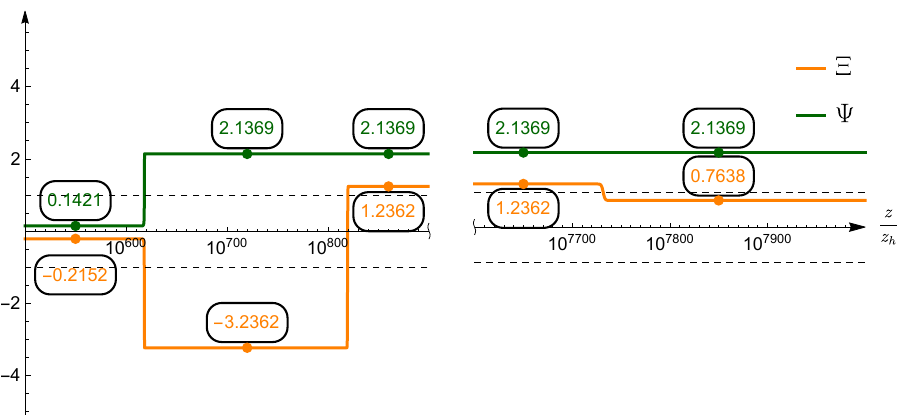}
    \caption{Kasner geometry inside the holographic superconductor with coexistence of s-wave order and p-wave order. The temperature in this case is chosen to be $T/T_c=0.648941$. In this specific temperature, there are multiple rounds of generalized Kasner inversion. \textbf{Left Panel:}Kasner transition behavior after first inversion  \textbf{Right Panel:} Kasner transition behavior after second inversion.}
    \label{kasner fig}
\end{figure*}
The above analysis fits the numerical results very well. Fig.~\ref{kasner fig} gives a numerical example where after Kasner transition and reflection, next-round generalized Kasner inversion behavior happens.
%The above analysis fits the numerical results very well. Fig.~\ref{kasner fig} gives a numerical example where after Kasner transition, reflection and generalized  inversion, stable Kasner epoch happens.
We show the internal structure of the black hole when the temperature is $T/T_c=0.648941$. The orange and green lines represent the parameters $\beta$ and $\gamma$ respectively. The two dotted lines with a value of $1$ and $-1$ distinguish the different Kasner alternations. 
%After passing through the collapse and oscillation near the event horizon, the system enter the Kasner epoch. 
At the beginning of the Kasner epoch, $\beta^2+\gamma^2<1$ satisfies the generalized Kasner inversion condition. Both $\beta$ and $\gamma$ increase proportionally. 
Then, $|\beta|>1$ which leads to Kasner transition and Kasner reflection respectively. During the transition and reflection period, the dynamics is irrelevant to the scalar hair, thus $\gamma$ remains unchanged until the next generalized Kasner inversion period occurs.
The second generalized Kasner inversion occurs near $z/z_h=10^{618}$. It is notable that this generalized inversion makes $|\gamma|>1$. Therefore, when $|\beta|<1$ after the Kasner transition and reflection, the system will settle down to a stable Kasner epoch towards the black hole singularity. The above analysis was well verified at $z/z_h=10^{7730}$.

\section{Conclusion and Discussion:}\label{sec5}
In this work, we investigate the interior structure of holographic multi-band superconductor model. We generalize previous investigations to the $s+p$ case where there is coexistence of scalar and vector order parameters. For single p-wave order parameter, it has been shown that the spacetime structure near the singularity is of chaotic type with infinite transitions among Kasner epochs \cite{Cai:2021obq,Cai:2024ltu}. However, with the addition of scalar degrees of freedom, the chaotic singularity vanishes which transits to a stable singularity when scalar hair appear. Therefore, black hole singularity structure can be a sensitive probe to reflect the coexistence region of boundary superconducting system. 

The transition of singularity structure also matches the results obtained by the phase space method. Through Hamiltonian analysis\cite{Damour:2002et, Henneaux:2022ijt}, it is well established that the evolution of spacetime metric can be mapped to the motion of a massless free particle moving in the hyperbolic space.
%We begin with the (d+1)-dimensional Einstein-Maxwell-vector theory \cite{Cai:2024ltu}. By selecting an appropriate metric coordinate, the authors demonstrate that the chaotic Kasner epoch within the black hole interior can be described by a independent parameter. Through Hamiltonian analysis\cite{Damour:2002et, Henneaux:2022ijt}, the spacetime structure can be mapped to the motion of a massless free particle moving in the d-dimensional hyperbolic space. In this framework, the non-diagonal Kasner metrics of the hyperbolic space form (d-1) symmetry walls and one electric wall, which confines the particle within a finite volume, causing it to undergo repeated collisions. 
Building upon this framework,  Ref.\cite{Henneaux:2022ijt} shows that the singularity structures are vastly distinct between black hole with massive and massless vector field. The reason is that the massive vector field will have an additional longitude mode which makes the structure of billiard table changes from finite volume case to infinite volume case. As longitude mode can be equivalently treated as a scalar degree of freedom, the scalar field in our case plays the role as the longitude mode in Ref.\cite{Henneaux:2022ijt}. It extends the hyperbolic billiard table by an additional dimension which the walls can not bound. Consequently, after finite number of collisions, the particle escapes into a direction devoid of walls which leads to the stable Kasner epoch. In this sense, our work constitutes a real space demonstration of the above phenomenon originally raised in Ref.\cite{Henneaux:2022ijt}. 

Many further directions need to be pursued. Firstly, based on the background we found, we can calculate some holographic observables which can reflect the singularity structure, such as thermal a-function \cite{Caceres:2022smh,Caceres:2022hei}, operator corelation function \cite{David:2022nfn,Grinberg:2020fdj,Fidkowski:2003nf},entanglement \cite{Anegawa:2024kdj} and complexity \cite{Jorstad:2023kmq}. Furthermore, by taking into account quantum corrections  for the near singularity region, there are additional novel spacetime structures which is called "Kasner eons" in Ref.\cite{Bueno:2024fzg,Caceres:2024edr,Bueno:2024qhh}. It is interesting to investigate this phenomenon in our model.

\section*{Acknowledgements}
We are grateful for the useful discussions with our group members. This work is supported by the National Natural Science Foundation of China (NSFC) under Grant Nos.12405066, 12175105 and 11965013. YSA is also supported by the Natural Science Foundation of Jiangsu Province under Grant No. BK20241376 and Fundamental Research Funds for the Central Universities. 


\appendix
%\section{Proof of no-Cauchy horizon theorem by conserved charge method:}
%As the model has scaling symmetry, there is a conserved charge associated to this symmetry by Noether theorem. The conserved charge reads
%\begin{equation}
%   Q(z)=e^{\chi/2}[z^{-2}(fe^{-\chi})'-\phi\phi'].
%\end{equation}
%It can be directly verified that $Q'(z)=0$ by using the equation of motion from Eq.(\ref{eom3}) to Eq.(\ref{eom5}). Assuming the black hole has two horizons at $z_{h}$ and $z_{i}$, thus the blackening factor $f(z)$ will satisfy the condition 
%\begin{equation}\label{rho}
%\begin{aligned}
%    &f(z_{h})=0,\quad f'(z_{h})<0, \\&
%    f(z_{i})=0, \quad f'(z_{i})>0.
%\end{aligned}
%\end{equation}
%There are also constraints on the gauge field by the regularity of equation of motion
%\begin{equation}\label{reg}
%    \phi(z_{h})=\phi(z_{i})=0.
%\end{equation}
%By using the conditions (\ref{rho}) and (\ref{reg}), we can find 
%\begin{equation}
%    Q(z_{h})=Q(z_{i}) \to \frac{f'(z_{h})}{z_{h}^{2}}e^{-\chi(z_{h})/2}=\frac{f'(z_{i})}{z_{i}^{2}}e^{-\chi(z_{i})/2}.
%\end{equation}
%This can not be true as the blackening factor must obey Eq.(\ref{rho}) as there are two horizons. Thus the assumption that the black hole has two horizons is wrong and there is only one horizon for black hole with scalar hair. Note that his proof is also valid for $k=+1$ case, however we only focus on $k=0$ case in our paper as our holographic superconductor is in $k=0$ case. 

\bibliographystyle{elsarticle-num} 
\biboptions{sort&compress}
\bibliography{msc}

\end{document}